UDC 519.635

# Towards identification of explicit solutions to overdetermined systems of differential equations

Maxim Zaytsev[1], V'yacheslav Akkerman[2]

[1] Nuclear Safety Institute of Russian Academy of Sciences

115191 Moscow, ul. Bolshaya Tulskaya, 52, Russia

[2] West Virginia University Morgantown, WV 26506-6106, USA

## Abstract

The authors proposed a general way to find particular solutions for overdetermined systems of PDEs previously, where the number of equations is greater than the number of unknown functions. In this paper, we propose an algorithm for finding solutions for overdetermined PDE systems, where we use a method for finding an explicit solution for overdetermined algebraic (polynomial) equations. Using this algorithm, the solution of some overdetermined PDE systems can be obtained in explicit form. The main difficulty of this algorithm is the huge number of polynomial equations that arise, which need to be investigated and solved numerically or explicitly. For example, the overdetermined hydrodynamic equations obtained earlier by the authors give a minimum of 10 million such equations. However, if they are solved explicitly, then we can write out the solution of the hydrodynamic equations in a general form, which is of great scientific interest.

## 1. Introduction

Partial differential equations (equations of mathematical physics) are often found in various fields of mathematics, physics, mechanics, chemistry, biology, and in numerous applications [1, 2]. The authors proposed a general way to reduce the dimension for arbitrary systems of partial differential equations (PDE), which allows reducing the PDE systems in volume to systems on the surface [3-6]. From the point of view of numerical methods, reduction is beneficial in the sense that it is not necessary to solve differential equations in the whole space. It is required to supplement the original PDE system with additional constraint equations and make transformations. On the basis of this idea, the authors of [5, 6] also proposed a method for finding particular solutions for overdetermined PDE systems, where the number of equations is greater than the number of unknown functions, which are very important for numerical calculations. In this method, finding solutions reduces to solving systems of ordinary implicit equations. In the articles of the authors [3-7], overdetermined equations of hydrodynamics are given, as well as methods for overriding any systems of PDE. In this paper, we propose an algorithm for finding solutions for overdetermined PDE systems, where we use a method for finding an explicit solution for overdetermined algebraic equations. Using this algorithm, the solution of some overdetermined PDE systems can be obtained in explicit form.

## 2. Method Description

It is required to find solutions for an overdetermined system of first-order $p+n$ partial differential equations with respect to unknowns $S_v(\mathbf{x})$, $v=1...p$, $\mathbf{x}=(x_1,...x_m)$

$$H_k\left(...\frac{\partial S_v}{\partial \mathbf{x}},...S_v,...\mathbf{x}\right)=0,\ v=1...p,\ k=1...(p+n), \tag{1}$$

where $H_k\left(...\frac{\partial S_v}{\partial \mathbf{x}},...S_v,...\mathbf{x}\right)$, $k=1...(p+n)$ smooth enough functions of their arguments $\partial S_v/\partial \mathbf{x}$, $S_v$, $\mathbf{x}=(x_1,...x_m)\in\mathbb{R}^m$, $v=1...p$. We apply the method described in articles [5, 6]. We consider the following system of equations of the form

$$P_\alpha(...Q_\beta,...\mathbf{x})=\frac{\partial^{(i_1+...+i_m)}}{\partial x_1^{i_1}...\partial x_m^{i_m}}\left[H_k\left(...\frac{\partial S_v}{\partial \mathbf{x}},...S_v,...\mathbf{x}\right)\right]=0,\ v=1...p,\ k=1...(p+n) \tag{2}$$

with respect to unknowns

$$Q_\beta=\frac{\partial^{(j_1+...+j_m)}S_v}{\partial x_1^{j_1}...\partial x_m^{j_m}},\ v=1...p. \tag{3}$$

Here, $\alpha = \alpha(k, i_1, ...i_m) = 1...N_H$ , $\beta = \beta(v, j_1, ...j_m) = 1...N_S$ the functions of indices (multi-indices) are such that

$$Q_{\beta(v,0,...0)} = S_v \,,\ v = 1...p \,, \tag{4}$$

$$P_{\alpha(k,0,...0)}\left(...Q_\beta,...\mathbf{x}\right) = H_k\left(Q_{\beta(v,1,...0)},...Q_{\beta(v,0,...1)},...Q_{\beta(v,0,...0)},...\mathbf{x}\right) \,, k = 1...(p+n) \,, \tag{5}$$

$$i_1 = 0...(N_1 - 1),\ i_2 = 0...(N_2 - 1),\ \dots\ i_m = 0...(N_m - 1), \tag{6}$$

$$j_1 = 0...N_1,\ j_2 = 0...N_2,\ \dots\ j_m = 0...N_m \,. \tag{7}$$

We also have

$$N_H = (p+n) N_1 N_2 ... N_m , \tag{8}$$

$$N_S = p \cdot (N_1 + 1)(N_2 + 1)...(N_m + 1) \,. \tag{9}$$

Let

$$\alpha = \alpha(k, i_1, ...i_m) = k + i_1 (p+n) + i_2 (p+n) N_1 + ... + i_m (p+n) N_1 ... N_{m-1} \,,$$

$$\beta = \beta(v, j_1, ...j_m) = v + j_1 p + j_2 p (N_1 + 1) + ... + j_m p (N_1 + 1)...(N_{m-1} + 1) \,.$$

Consider the matrix

$$A_{\alpha\beta} = \left( \frac{\partial P_\alpha}{\partial Q_\beta} \right),\ \alpha = 1...N_H \,,\ \beta = 1...N_S \,. \tag{10}$$

Let its rank be equal to $N_S^{real}$ the number of unknowns $Q_\beta$ actually presented in equations (2)

$$N_S^{real} \le N_H \frac{p}{(p+n)} \left( 1 + \sum_{l=1}^{m} \frac{1}{N_l} \right). \tag{11}$$

Consider an **extended** overdetermined system of implicit equations of the form

$$P_{\alpha^*}\left(...Q_{\beta^*},...\mathbf{x}\right) = \frac{\partial^{(i_1+...+i_m)}}{\partial x_1^{i_1} ... \partial x_m^{i_m}} \left[ H_k \left( ... \frac{\partial S_v}{\partial \mathbf{x}}, ...S_v, ...\mathbf{x} \right) \right] = 0 \,,\ v = 1...p \,,\ k = 1...(p+n) \tag{12}$$

with respect to unknowns

$$Q_{\beta^*} = \frac{\partial^{(j_1+...+j_m)} S_v}{\partial x_1^{j_1} ... \partial x_m^{j_m}} \,,\ v = 1...p \,. \tag{13}$$

But functions from multi-indices $\alpha^* = \alpha^*(k, i_1, ...i_m) = 1...N_H^w$, $\beta^* = \beta^*(v, j_1, ...j_m) = 1...N_S^w$, contrary to (6), (7) let be determined from the conditions:

$$i_1 = 0...N_1,\ i_2 = 0...N_2,\ \dots\ i_m = 0...N_m,\ k = 1...(p+n), \tag{14}$$

$$j_1 = 0...(N_1 + 1),\ j_2 = 0...(N_2 + 1),\ \dots\ j_m = 0...(N_m + 1),\ v = 1...p \,. \tag{15}$$

Let

$\alpha^* = \alpha^*(k, i_1, \dots i_m) = k + i_1(p+n) + i_2(p+n)(N_1+1) + \dots + i_m(p+n)(N_1+1)\dots(N_{m-1}+1)$,

$\beta^* = \beta^*(v, j_1, \dots j_m) = v + j_1 p + j_2 p(N_1+2) + \dots + j_m p(N_1+2)\dots(N_{m-1}+2)$.

We also have

$$N_H^w = (p+n)(N_1+1)(N_2+1)\dots(N_m+1), \tag{16}$$

$$N_S^w = p \cdot (N_1+2)(N_2+2)\dots(N_m+2). \tag{17}$$

Differentiating expression (12) with respect to the variable $x_s$, $s = 1\dots m$, we find that

$$P_{\tilde{\alpha}}\left(\dots Q_{\beta^*}, \dots \mathbf{x}\right) = \sum_{v=1}^{p} \sum_{j_1=0}^{N_1+1} \sum_{j_2=0}^{N_2+1} \dots \sum_{j_m=0}^{N_m+1} \frac{\partial P_{\alpha^*}}{\partial Q_{\beta^*}} Q_{\tilde{\beta}} + \frac{\partial P_{\alpha^*}\left(\dots Q_{\beta^*}, \dots \mathbf{x}\right)}{\partial x_s} = 0, \tag{18}$$

where

$$\tilde{\alpha} = \alpha^*\left(k, i_1, \dots \left(i_s+1\right) \dots i_m\right),\ \tilde{\beta} = \beta^*\left(v, j_1, \dots \left(j_s+1\right) \dots j_m\right),\ s = 1\dots m.$$

Here, indices are taken such that, $i_s \le N_s - 1$, $j_s \le N_s$.

Any of equations (12) (as well as (18)) with an index $\alpha^* = \alpha^*(k, i_1, \dots i_m)$ contains unknowns of the form (13) with indices $\beta^* = \beta^*(v, j_1, \dots j_m)$, where $0 \le j_1 \le i_1 + 1$, … $0 \le j_m \le i_m + 1$ and $0 \le j_1 + \dots + j_m \le i_1 + \dots + i_m + 1$. Obviously

$$\frac{\partial P_{\alpha^*\left(k, i_1, \dots i_s = (N_s - 1) \dots i_m\right)}}{\partial Q_{\beta^*\left(v, j_1, \dots j_s = N_s + 1, \dots j_m\right)}} = 0, \tag{19}$$

that is, there are no terms with an index $\tilde{\beta} = \beta^*\left(v, j_1, \dots \left(j_s+1\right) = N_s + 2 \dots j_m\right)$. Therefore, the recurrence relation (18) is correct.

**Solution method.** We recursively determine implicit equations from (18) using (4), (5) and solve them. We calculate the rank (10) on these solutions. Calculate manually the number of variables $N_S^{real}$ (11)! and compare with the rank of the matrix (10). If they coincide, then we have a solution to system (2) (see (4)). Obviously, we are only interested in the case: $N_H \ge N_S$.

We also have the following estimates [5, 6]

$$N_H \ge (p+n)(mp/n)^m, \tag{20}$$

where the minimum is realized at $N_1 \approx N_2 \approx \dots \approx N_m \approx mp/n$.

The system of equations (12), if it is sufficiently "good", can be somewhat simplified.

Consider an overdetermined system of $p+n$ partial differential equations of the first order with respect to unknowns $S_v(\mathbf{x})$, $v = 1\dots p$, $\mathbf{x} = (x_1, \dots x_m)$ (1). Suppose that $n \ge (m-1)p$. In the case of (ODE) $m=1$, this is always the case. Consider some $(p+n)$ equations with a multi-index $\alpha^* = \alpha^*(k, i_1, ...i_m)$ from the system of equations (12) with fixed indices $(i_1, ...i_m)$, but $k = 1...(p+n)$. By the formula (18), they can be written as

$$P_{\alpha^*}\left(...Q_{\beta^*}, ...\mathbf{x}\right) = \sum_{v=1}^{p}\sum_{s=1}^{m}\frac{\partial P_{\tilde{\alpha}_0}}{\partial Q_{\beta_s^0}} Q_{\tilde{\beta}_s} + ... = 0\,, \tag{21}$$

where $\alpha^* = \alpha^*(k, i_1, ...i_m)$, $\tilde{\alpha}_0 = \alpha^*(k, 0, ...0)$, $\beta_s^0 = \beta^*\left(v, 0, ...\underset{s}{1}...0\right)$, $\tilde{\beta}_s = \beta^*\left(v, i_1, ...(i_s+1)...i_m\right)$, $s = 1...m$, $v = 1...p$, $k = 1...(p+n)$.

Expressions $P_{\alpha^*}\left(...Q_{\beta^*}, ...\mathbf{x}\right)$ contain unknowns $Q_{\beta^*}$, in which $0 \le j_1 + ... + j_m \le i_1 + ... + i_m + 1$ and $0 \le j_1 \le i_1 + 1$, $\dots$ $0 \le j_m \le i_m + 1$. Unknowns $Q_{\beta^*}$, in which $j_1 + ... + j_m = i_1 + ... + i_m + 1$, enter into equations (21) linearly. Obviously, their number is equal to $mp$. Therefore, if the corresponding determinant is not equal to zero, all of them can be expressed explicitly from the system of $(p+n) \ge mp$ equations (21), which is linear with respect to them.

Thus, it can be argued that in our case any unknown $Q_{\beta^*}$, where $\beta^* = \beta^*(v, j_1, ...j_m)$, $v = 1...p$ is expressed explicitly via unknowns $Q_{\beta^0}$, where $\beta^0 = \beta^0(v, j_1^0, ...j_m^0)$, $v = 1...p$ and $0 \le j_1^0 + ... + j_m^0 < j_1 + ... + j_m$, $0 \le j_1^0 \le j_1 + 1$, $\dots$ $0 \le j_m^0 \le j_m + 1$ therefore, via $Q_{\beta_s^0}$, where $\beta_s^0 = \beta^*\left(v, 0, ...\underset{s}{1}...0\right)$ and $v = 1...p$, $s = 1...m$ or $Q_{\beta^*(v,0,...0)} = S_v$, $v = 1...p$. As a result, instead of (12), we obtain an overdetermined system of implicit equations with respect to $Q_{\beta^*(v,0,...0)} = S_v$, $v = 1...p$.

## 3. A method for solving system of overdetermined algebraic equations by their reduction

Equations (2) and (12) with respect to unknowns $Q_\beta$, $j_1 + ... + j_m \ge 2$, $v = 1...p$ are algebraic. There are various methods for solving polynomial equations. In particular, the far nontrivial Bukhberger algorithm [8] is widespread. In our case, we can apply the following simple method to the solution of systems of overdetermined algebraic equations.

Consider an overdetermined system of two polynomials of order $n$

$$\sum_{i=0}^{n} a_i x^i = 0, \tag{22}$$

$$\sum_{i=0}^{n} b_i x^i = 0, \tag{23}$$

where $a_i \in \mathbb{R}$, $b_i \in \mathbb{R}$, $i = 0...n$, $a_n \neq 0$. It follows from (22) that

$$a_n x^n = -\sum_{i=0}^{n-1} a_i x^i, \; x^n = \frac{1}{a_n}\left(-\sum_{i=0}^{n-1} a_i x^i\right). \tag{24}$$

We substitute (24) into (23). We have

$$b_n x^n + \sum_{i=0}^{n-1} b_i x^i = 0, \; b_n \frac{1}{a_n}\left(-\sum_{i=0}^{n-1} a_i x^i\right) + \sum_{i=0}^{n-1} b_i x^i = 0. \tag{25}$$

As a result, we obtain a polynomial of the form:

$$\sum_{i=0}^{n-1} c_i x^i = 0, \tag{26}$$

where $c_j = b_j - \frac{b_n}{a_n} a_j$, $j = 0...(n-1)$. Let

$$c_{n-1} \neq 0. \tag{27}$$

We multiply both sides of (26) by $x$ and substitute instead of $x^n$ its expression from (24):

$$\sum_{i=0}^{n-1} c_i x^{i+1} = 0, \tag{28}$$

$$c_{n-1} x^n + \sum_{i=0}^{n-2} c_i x^{i+1} = 0, \tag{29}$$

$$\frac{c_{n-1}}{a_n}\left(-\sum_{i=0}^{n-1} a_i x^i\right) + \sum_{i=0}^{n-2} c_i x^{i+1} = 0. \tag{30}$$

As a result, we obtain a polynomial of the form:

$$\sum_{i=0}^{n-1} d_i x^i = 0, \tag{31}$$

where $d_j = c_{j-1} - \frac{c_{n-1}}{a_n} a_j$, $j = 0...(n-1)$, $c_{-1} = 0$.

Thus, we have obtained two polynomials (26) and (31) of order no more than $n-1$. We show that the sets of solutions of the system of equations (22), (23) and system (26), (31) coincide, under condition (27). We have shown that (26), (31) follows from (22), (23). Let us prove the opposite. Let $x$ be some solution (26), (31). Multiply both sides of (26) by $x$. Then equation (29) holds. On the other hand, equation (30) is satisfied, since this is another form of notation (31). Subtract equation (30) term by term from (29). Then we find that

$$c_{n-1}\left[x^n+\frac{1}{a_n}\left(\sum_{i=0}^{n-1}a_i x^i\right)\right]=0. \tag{32}$$

Therefore, by virtue of condition (27), equation (22) holds. Equation (23) then is a consequence of (22) and (26). Q.E.D.

**Example**. Consider an overdetermined system of two quadratic equations

$$ax^2+bx+c=0, \tag{33}$$

$$px^2+qx+r=0. \tag{34}$$

Let $a\neq 0$. It follows from (33) that

$$x^2=-\frac{b}{a}x-\frac{c}{a}. \tag{35}$$

We substitute (35) into (34). Then

$$\left(q-\frac{pb}{a}\right)x+\left(r-\frac{pc}{a}\right)=0. \tag{36}$$

Multiply (36) by $x$ and substitute the expression from (35) instead of $x^2$. Consequently,

$$\left(r-\frac{pc}{a}-\frac{qb}{a}+\frac{pb^2}{a^2}\right)x+\left(\frac{pbc}{a^2}-\frac{qc}{a}\right)=0. \tag{37}$$

The system of equations (36), (37) is equivalent to the system (33), (34) under the condition

$$\left(q-\frac{pb}{a}\right)\neq 0. \tag{38}$$

System (36), (37) has a solution

$$x=-\frac{\left(r-\frac{pc}{a}\right)}{\left(q-\frac{pb}{a}\right)} \tag{39}$$

provided

$$\begin{vmatrix} \left(q-\frac{pb}{a}\right) & \left(r-\frac{pc}{a}\right) \\ \left(r-\frac{pc}{a}-\frac{qb}{a}+\frac{pb^2}{a^2}\right) & \left(\frac{pbc}{a^2}-\frac{qc}{a}\right) \end{vmatrix}=0. \tag{40}$$

If condition (40) is not satisfied, but (38) is true, then system (33), (34) has no solutions.

In the case of an overdetermined system of more than two polynomials, one can prove a similar statement and find the condition when the original system is equivalent to a system of the same number of polynomials, but of an order less than that of the original system.

Consider the general case of an overdetermined system of $(m+1)$ algebraic equations in $m$ variables

$$\sum_{i_1=0}^{n_1}\sum_{i_2=0}^{n_2}...\sum_{i_m=0}^{n_m} a^j_{(i_1,i_2,...i_m)} \left(x_1\right)^{i_1}\cdot\left(x_2\right)^{i_2}\cdots\left(x_m\right)^{i_m}=0, \tag{41}$$

where $a^j_{(i_1,i_2,...i_m)}\in\mathbb{R}$, $i_1=0...n_1$, $\ldots$ $i_m=0...n_m$, $j=1...(m+1)$. System (41) can be transformed to the form

$$\sum_{i_m=0}^{n_m} A^j_{i_m}\left(x_m\right)^{i_m}=0, \tag{42}$$

where

$$A^j_{i_m}=\sum_{i_1=0}^{n_1}\sum_{i_2=0}^{n_2}...\sum_{i_{m-1}=0}^{n_{m-1}} a^j_{(i_1,i_2,...i_m)}\left(x_1\right)^{i_1}\cdot\left(x_2\right)^{i_2}\cdots\left(x_{m-1}\right)^{i_{m-1}}, \tag{43}$$

$$i_1=0...n_1,\ \ldots\ i_m=0...n_m,\ j=1...(m+1).$$

The system of equations (42) can be considered as an overdetermined system of $(m+1)$ algebraic equations with respect to a variable $x_m$, but with variable coefficients (43). Under certain conditions, it can be equivalently reduced sequentially to a system of $(m+1)$ linear equations with respect to the variable $x_m$ and find its solution in the form of a fractional rational expression of type (39**)** and of the coefficients (43). The conditions for the existence of this solution in this form will be some $m$ fractional rational expressions of type (40), composed of coefficients (43). Thus, in fact, we get an overdetermined system of $m$ algebraic equations of the form (41), where the number of variables $m-1$ is reduced by one, and so on.

Thus, it is possible to exclude unknowns $Q_{\beta^*}$, $j_1+...+j_m\geq 2$, $v=1...p$ from equations (12) under certain conditions. As a result, only an overdetermined system of implicit equations from unknowns $Q_{\beta^0_s}$, where $\beta^0_s=\beta^*\left(v,0,\underset{s}{...1}...0\right)$, $v=1...p$, $s=1...m$ and $Q_{\beta^*(v,0,...0)}=S_v$, $v=1...p$, remains, which is much easier to solve than the original one (12).

## 4. Conclusion

In this article, we propose an algorithm for finding solutions for overdetermined PDE systems based on the authors' works [5, 6]. The main difficulty of this algorithm is the huge number of polynomial equations that appear in it, which need to be investigated and solved numerically or explicitly. The authors of [3, 4] present overdetermined equations of hydrodynamics, which give a minimum of 10 million such equations. However, if they are solved explicitly, then the solution of the equations of hydrodynamics can be written out, which is of great interest. In this paper, we also propose an easy-to-describe algorithm for finding an explicit solution for systems of overdetermined algebraic equations if for these systems some requirements are fulfilled related to the steps of this algorithm. In the case of its successful implementation, this will be a new and useful tool in the study of the PDE systems.